\documentclass[graybox,vecphys]{svmult}

\usepackage{helvet} % selects Helvetica as sans-serif font
\usepackage{courier} % selects Courier as typewriter font
\usepackage{type1cm} % activate if the above 3 fonts are
\usepackage{makeidx} % allows index generation
\usepackage{graphicx} % standard LaTeX graphics tool
\usepackage{multicol} % used for the two-column index
\usepackage[bottom]{footmisc}% places footnotes at page bottom

\usepackage{amsmath,amssymb,tikz,epic,eepic}%
\usepackage{hyperref}%
\hypersetup{colorlinks=true,%
  linkcolor=blue,%
  citecolor=magenta,%
  urlcolor=black,%
  bookmarksnumbered=true,%
  bookmarkstype=toc,%
  bookmarksopen=false,%
  pdftitle={Conformal Ward-Takahashi Identity at Finite Temperature},%
  pdfauthor={Satoshi Ohya}}%

\begin{document}

\title*{Conformal Ward--Takahashi Identity at Finite Temperature}%
\author{Satoshi Ohya}%
\institute{Institute of Quantum Science, Nihon University,
  Kanda-Surugadai 1-8-14, Chiyoda, Tokyo 101-8308, Japan.
  \email{ohya@phys.cst.nihon-u.ac.jp}}

\maketitle

\abstract{We study conformal Ward--Takahashi identities for two-point
  functions in $d(\geq3)$-dimensional finite-temperature conformal
  field theory. We first show that the conformal Ward--Takahashi
  identities can be translated into the intertwining relations of
  conformal algebra $\mathfrak{so}(2,d)$. We then show that, at finite
  temperature, the intertwining relations can be translated into the
  recurrence relations for two-point functions in complex momentum
  space. By solving these recurrence relations, we find the
  momentum-space two-point functions that satisfy the
  Kubo--Martin--Schwinger thermal equilibrium condition.}

\section{Introduction}
\label{sec:1}
It is widely believed that conformal symmetry is always broken at
finite temperature. This comes from the naive argument that
finite-temperature field theory necessarily contains one particular
scale---the temperature---and hence must break scale and conformal
invariance. Contrary to this popular belief, however, finite
temperature and conformal invariance can in fact be compatible with
each other: If conformal field theory (CFT) is thermalized via the
\textit{Unruh effect}, conformal symmetry remains intact even at
finite temperature. The purpose of this paper is to report our recent
work on this subject \cite{Ohya:2016gto} and to see how the conformal
symmetry determines finite-temperature two-point functions in momentum
space. The key is the \textit{intertwining relations} of conformal
algebra $\mathfrak{so}(2,d)$
\cite{Dobrev:1977qv,Fradkin:1978pp,Koller:1974ut,Todorov:1978rf},
which follow from the conformal Ward--Takahashi identities for
two-point functions. We shall show that, at finite temperature, the
intertwining relations are recast into the recurrence relations in
complex momentum space. These recurrence relations can be regarded as
the conformal Ward--Takahashi identities at finite temperature, from
which we can deduce the possible forms of momentum-space two-point
functions.

The rest of the paper is organized as follows: In Section \ref{sec:2}
we first introduce the intertwining operator, which is defined as an
integral transform whose kernel is the two-point function. We then
discuss that the conformal Ward--Takahashi identities are rewritten as
the intertwining relations. In Section \ref{sec:3} we introduce the
$d(\geq3)$-dimensional Rindler wedge, light-cone, and diamond, all of
which are subspaces of Minkowski spacetime and conformal to
$\mathbb{H}^{1}\times\mathbb{H}^{d-1}$. These subspaces are the whole
universes of our finite-temperature CFT and possess the global
timelike conformal Killing vectors associated with the subgroup
$SO(1,1)\subset SO(2,d)$. In Section \ref{sec:4} we study the
intertwining relations in the basis in which the $SO(1,1)$ generator
becomes diagonal. We shall see that in this basis the intertwining
relations reduce to the recurrence relations for momentum-space
two-point functions. We also give two minimal solutions that
correspond to the positive- and negative-frequency two-point Wightman
functions and satisfy the Kubo--Martin--Schwinger (KMS) thermal
equilibrium condition.

Throughout the paper we work with the metric signature
$(-,+,\cdots,+)$.

\section{From Conformal Ward--Takahashi Identities to Intertwining
  Relations}
\label{sec:2}
To begin with, let us consider a scalar primary operator
$\mathcal{O}_{\Delta}(x)$ of scaling dimension $\Delta$. Let
$g\in SO(2,d)$ be an element of the conformal group and
$x\mapsto x_{g}$ be the associated conformal transformation. Then the
scalar primary operator transforms as follows:
\begin{align}
  U(g)\mathcal{O}_{\Delta}(x)U^{-1}(g)=\left|\frac{\partial x_{g}}{\partial x}\right|^{\Delta/d}\mathcal{O}_{\Delta}(x_{g}),\label{eq:1}
\end{align}
where $U$ is a unitary representation of the conformal group and
$|\partial x_{g}/\partial x|$ stands for the Jacobian of the conformal
transformation.

Let us next consider a two-point function $G_{\Delta}(x,y)$ of
$\mathcal{O}_{\Delta}$. For example, one may consider this to be the
positive- or negative-frequency two-point Wightman functions,
$\langle0|\mathcal{O}_{\Delta}(x)\mathcal{O}_{\Delta}^{\dagger}(y)|0\rangle$
or
$\langle0|\mathcal{O}_{\Delta}^{\dagger}(y)\mathcal{O}_{\Delta}(x)|0\rangle$,
where $|0\rangle$ stands for the conformally-invariant vacuum state
that satisfies $U(g)|0\rangle=|0\rangle$ for any $g\in SO(2,d)$. Then
$G_{\Delta}(x,y)$ satisfies the following identity:
\begin{equation}
  G_{\Delta}(x,y)=\left|\frac{\partial x_{g}}{\partial x}\right|^{\Delta/d}\left|\frac{\partial y_{g}}{\partial y}\right|^{\Delta/d}G_{\Delta}(x_{g},y_{g}).\label{eq:2}
\end{equation}
As is well-known, this identity---the finite form of conformal
Ward--Takahashi identity---fully determines the possible forms of
two-point functions. For example, up to the $i\epsilon$ prescription
the Wightman functions must be of the form
$G_{\Delta}(x,y)\propto[(x-y)^{2}]^{-\Delta}$.

Now, let us consider another scalar primary operator
$\mathcal{O}_{d-\Delta}(x)$ of scaling dimension $d-\Delta$. Once
$\mathcal{O}_{d-\Delta}(x)$ and $G_{\Delta}(x,y)$ are given, we can
define an operator $G_{\Delta}$ through the following integral
transform:
\begin{equation}
  G_{\Delta}: \mathcal{O}_{d-\Delta}(x)\mapsto(G_{\Delta}\mathcal{O}_{d-\Delta})(x):=\int\!d^{d}y\,G_{\Delta}(x,y)\mathcal{O}_{d-\Delta}(y).\label{eq:3}
\end{equation}
It is easy to check that thus defined operator
$(G_{\Delta}\mathcal{O}_{d-\Delta})(x)$ satisfies the transformation
law \eqref{eq:1} and hence is a primary operator of scaling dimension
$\Delta$. Conversely, one can start from $\mathcal{O}_{\Delta}(x)$ and
$G_{d-\Delta}(x,y)$ and then define an operator $G_{d-\Delta}$ through
the integral
$(G_{d-\Delta}\mathcal{O}_{\Delta})(x):=\int
d^{d}y\,G_{d-\Delta}(x,y)\mathcal{O}_{\Delta}(y)$. In this case
$(G_{d-\Delta}\mathcal{O}_{\Delta})(x)$ becomes a primary operator of
scaling dimension $d-\Delta$. In short, $G_{\alpha}$ is a map from one
primary operator to another, where $\alpha\in\{\Delta,d-\Delta\}$. In
the literature \cite{Ferrara:1972uq}
$(G_{\alpha}\mathcal{O}_{d-\alpha})(x)$ is called the shadow operator
of $\mathcal{O}_{d-\alpha}(x)$.

Let us now turn to the infinitesimal conformal invariance. If
$g\in SO(2,d)$ is infinitesimally close to the identity element,
\eqref{eq:1} is recast into the following commutation relations:
\begin{equation}
  [J^{ab},\mathcal{O}_{\Delta}(x)]=-J^{ab}_{\Delta}(x,\partial_{x})\mathcal{O}_{\Delta}(x).\label{eq:4}
\end{equation}
Likewise, \eqref{eq:2} becomes the following identities (the
infinitesimal form of conformal Ward--Takahashi identities):
\begin{equation}
  \left(J^{ab}_{\Delta}(x,\partial_{x})+J^{ab}_{\Delta}(y,\partial_{y})\right)G_{\Delta}(x,y)=0.\label{eq:5}
\end{equation}
Here $J^{ab}=-J^{ba}$ ($a,b=0,1,\cdots,d+1$) are the generators of
$SO(2,d)$ and satisfy the following commutation relations of the Lie
algebra $\mathfrak{so}(2,d)$:
\begin{equation}
  [J^{ab},J^{cd}]=i(\eta^{ac}J^{bd}-\eta^{ad}J^{bc}-\eta^{bc}J^{ad}+\eta^{bd}J^{ac}),\label{eq:6}
\end{equation}
where $\eta_{ab}=\eta^{ab}=\mathrm{diag}(-1,+1,\cdots,+1,-1)$. On the
other hand, $J_{\Delta}^{ab}(x,\partial_{x})$ are the following
differential representations of $J^{ab}$:
\begin{equation}
  J^{ab}_{\Delta}(x,\partial_{x})
  =i\left(k^{\mu ab}(x)\partial_{\mu}+\frac{\Delta}{d}(\partial_{\mu}k^{\mu ab})(x)\right),\label{eq:7}
\end{equation}
where $k^{\mu ab}(x)=-k^{\mu ba}(x)$ are the conformal Killing vectors
given by
\begin{align}
  &k^{\mu\nu\lambda}(x)=\eta^{\mu\nu}x^{\lambda}-\eta^{\mu\lambda}x^{\nu},\qquad
    k^{\mu\nu d}(x)=\frac{\ell^{2}-x\cdot x}{2\ell}\eta^{\mu\nu}+\frac{x^{\mu}x^{\nu}}{\ell},\label{eq:8}\\
  &k^{\mu\nu,d+1}(x)=\frac{\ell^{2}+x\cdot x}{2\ell}\eta^{\mu\nu}-\frac{x^{\mu}x^{\nu}}{\ell},\qquad
    k^{\mu d,d+1}(x)=-x^{\mu}.\label{eq:9}
\end{align}
Here $\ell>0$ is an arbitrary reference length scale which needs to be
introduced to adjust the length dimensions of the equations. Note that
these vectors satisfy the conformal Killing equations
$\partial_{\mu}{k_{\nu}}^{ab}+\partial_{\nu}{k_{\mu}}^{ab}=\tfrac{2}{d}\eta_{\mu\nu}\partial_{\rho}k^{\rho
  ab}$.

Now, let $G_{\Delta}(x,y)$ satisfy the infinitesimal conformal
Ward--Takahashi identities \eqref{eq:5}. Then, upon integration by
parts one can prove the following identities:
\begin{equation}
  \int\!d^{d}y\,J^{ab}_{\Delta}(x,\partial_{x})G_{\Delta}(x,y)\mathcal{O}_{d-\Delta}(y)
  =\int\!d^{d}y\,G_{\Delta}(x,y)J^{ab}_{d-\Delta}(y,\partial_{y})\mathcal{O}_{d-\Delta}(y),\label{eq:10}
\end{equation}
or, more compactly,
\begin{equation}
  (J^{ab}_{\Delta}G_{\Delta}\mathcal{O}_{d-\Delta})(x)
  =(G_{\Delta}J^{ab}_{d-\Delta}\mathcal{O}_{d-\Delta})(x),\label{eq:11}
\end{equation}
where
$(J^{ab}_{\alpha}\mathcal{O}_{\alpha})(x):=J^{ab}_{\alpha}(x,\partial_{x})\mathcal{O}_{\alpha}(x)$,
$\alpha\in\{\Delta,d-\Delta\}$. Since this holds for arbitrary
$\mathcal{O}_{d-\Delta}$ we get the following operator identities:
\begin{equation}
  J_{\Delta}^{ab}G_{\Delta}=G_{\Delta}J_{d-\Delta}^{ab}.\label{eq:12}
\end{equation}
These are the intertwining relations, and in this respect $G_{\Delta}$
is called the intertwining operator. As is evident from the above
discussions the intertwining relations are essentially the same as the
conformal Ward--Takahashi identities. There is, however, a big
advantage of using \eqref{eq:12}: The operator identities
\eqref{eq:12} are basis independent and hence easy to manipulate in an
algebraic language. In the rest of the paper we shall apply the
intertwining relations to a certain (improper) basis for a
representation space of conformal algebra. In other words, we shall
apply \eqref{eq:12} to a mode function $f_{\alpha,p}(x)$ in terms of
which the operator $\mathcal{O}_{\alpha}(x)$ is expanded as
$\mathcal{O}_{\alpha}(x)=\int\!\!
\frac{d^{d}p}{(2\pi)^{d}}\,\Tilde{\mathcal{O}}_{\alpha}(p)f_{\alpha,p}(x)$. In
zero-temperature CFT such mode function is just the plane wave
$\mathrm{e}^{ip\cdot x}$. In this case the intertwining relations just
result in the well-known momentum-space conformal Ward--Takahashi
identities at zero temperature. In finite-temperature CFT thermalized
via the Unruh effect, on the other hand, $f_{\alpha,p}(x)$ becomes a
quite nontrivial function. In a more algebraic language,
$f_{\alpha,p}(x)$ is chosen to be an eigenfunction for the generator
of one-parameter subgroup $SO(1,1)\subset SO(2,d)$. Before going to
study the intertwining relations in the $SO(1,1)$ diagonal basis, let
us first recall the significance of $SO(1,1)$ for finite-temperature
CFT.

\section{Timelike Conformal Killing Vectors Associated with the
  Subgroup $SO(1,1)\subset SO(2,d)$}
\label{sec:3}
Let us start with the KMS condition \cite{Haag:1967sg}. The KMS
condition is a thermal equilibrium condition for quantum systems and
expressed as an analytic condition for positive- and
negative-frequency two-point Wightman functions $G^{+}(t)$ and
$G^{-}(t)$. It demands that (i) $G^{+}(t)$ ($G^{-}(t)$) should be an
analytic function on the strip $-\beta<\mathrm{Im}\,t<0$
($0<\mathrm{Im}\,t<\beta$); and (ii) $G^{+}(t)$ and $G^{-}(t)$ should
satisfy the following boundary conditions on the strips:
\begin{equation}
  G^{+}(t)=G^{-}(t+i\beta)~~\&~~G^{-}(t)=G^{+}(t-i\beta),\quad \forall t\in\mathbb{R},\label{eq:13}
\end{equation}
where $\beta=1/T$ is the inverse temperature. (For the moment we will
suppress the spatial coordinates.) The advantage of using the KMS
condition is that these analytic conditions remain valid even after
the thermodynamic limit. (Note that the extensive property of the free
energy $F=-(1/\beta)\log\mathrm{Tr}\,\mathrm{e}^{-\beta H}$ would
render the density matrix $\rho=\mathrm{e}^{-\beta(H-F)}$ ill-defined
in the thermodynamic limit.) For a full account of the KMS condition
we refer to \cite{Haag:1967sg,Haag:1996}.

\begin{figure}[t]
  \sidecaption[t]%
  \input{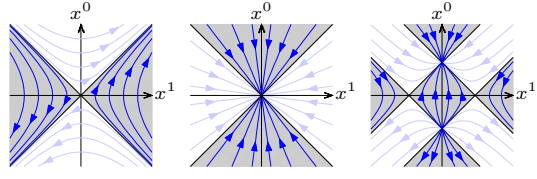}
  \caption{Timelike domains of the $SO(1,1)$ conformal Killing vectors
    \eqref{eq:20}--\eqref{eq:22} in the
    $(x^{0},x^{1})$-plane. Temporal flows are identified with thick
    blue curves.}
  \label{fig:1}
\end{figure}

Now, let us take a closer look at the boundary conditions
\eqref{eq:13}. These conditions are best understood in statistical
mechanics for finite degrees of freedom in a finite box. Let
$\mathcal{O}(t)=\mathrm{e}^{iHt}\mathcal{O}(0)\mathrm{e}^{-iHt}$ be a
Heisenberg operator. Then we have
\begin{align}
  \langle\mathcal{O}(t)\mathcal{O}^{\dagger}(t^{\prime})\rangle
  &=\frac{1}{Z}\mathrm{Tr}\left(\mathrm{e}^{-\beta H}\mathcal{O}(t)\mathcal{O}^{\dagger}(t^{\prime})\right)
    =\frac{1}{Z}\mathrm{Tr}\left(\mathrm{e}^{-\beta H}\mathcal{O}(t)\mathrm{e}^{\beta H}\mathrm{e}^{-\beta H}\mathcal{O}^{\dagger}(t^{\prime})\right)\nonumber\\
  &=\frac{1}{Z}\mathrm{Tr}\left(\mathrm{e}^{-\beta H}\mathcal{O}^{\dagger}(t^{\prime})\mathrm{e}^{-\beta H}\mathcal{O}(t)\mathrm{e}^{\beta H}\right)\nonumber\\
  &=\frac{1}{Z}\mathrm{Tr}\left(\mathrm{e}^{-\beta H}\mathcal{O}^{\dagger}(t^{\prime})\mathcal{O}(t+i\beta)\right)
    =\langle\mathcal{O}^{\dagger}(t^{\prime})\mathcal{O}(t+i\beta)\rangle,\label{eq:14}
\end{align}
where $Z=\mathrm{Tr}\,\mathrm{e}^{-\beta H}$ is the partition
function. The second line follows from the cyclic property of trace
and the last line the identity
$\mathrm{e}^{izH}\mathcal{O}(t)\mathrm{e}^{-izH}=\mathcal{O}(t+z)$
with $z=i\beta$. Setting $t^{\prime}=0$ we get the condition
$G^{+}(t)=G^{-}(t+i\beta)$. Likewise, one can prove
$G^{-}(t)=G^{+}(t-i\beta)$ in a similar manner.

The above discussion is based on the expectation value with respect to
the density matrix $\rho=\mathrm{e}^{-\beta H}/Z$. However, the
boundary conditions \eqref{eq:13} themselves can be formulated without
recourse to the density matrix. Suppose that there exist a state
$|\Omega\rangle$ and an antiunitary operator $J$ such that the
following identity holds:
\begin{equation}
  J\mathrm{e}^{-\frac{\beta}{2}H}\mathcal{O}(t)|\Omega\rangle=\mathcal{O}^{\dagger}(t)|\Omega\rangle,\label{eq:15}
\end{equation}
where $\mathcal{O}(t)$ is an arbitrary Heisenberg operator and $H$ is
assumed to satisfy $H|\Omega\rangle=0$. Once we have the identity
\eqref{eq:15}, we can prove that the Wightman functions with respect
to the state $|\Omega\rangle$ satisfy \eqref{eq:13}. Indeed, by using
the inner product notation $(\ast,\ast)$ we have (see also Chapter 5
of \cite{Takagi:1986kn})
\begin{align}
  \langle\Omega|\mathcal{O}(t)\mathcal{O}^{\dagger}(t^{\prime})|\Omega\rangle
  &=(|\Omega\rangle,\mathcal{O}(t)\mathcal{O}^{\dagger}(t^{\prime})|\Omega\rangle)
    =(\mathcal{O}^{\dagger}(t)|\Omega\rangle,\mathcal{O}^{\dagger}(t^{\prime})|\Omega\rangle)\nonumber\\
  &=(J\mathrm{e}^{-\frac{\beta}{2}H}\mathcal{O}(t)|\Omega\rangle,J\mathrm{e}^{-\frac{\beta}{2}H}\mathcal{O}(t^{\prime})|\Omega\rangle)\nonumber\\
  &=(\mathrm{e}^{-\frac{\beta}{2}H}\mathcal{O}(t^{\prime})|\Omega\rangle,\mathrm{e}^{-\frac{\beta}{2}H}\mathcal{O}(t)|\Omega\rangle)\nonumber\\
  &=(|\Omega\rangle,\mathcal{O}^{\dagger}(t^{\prime})\mathrm{e}^{-\beta H}\mathcal{O}(t)\mathrm{e}^{\beta H}\mathrm{e}^{-\beta H}|\Omega\rangle)\nonumber\\
  &=(|\Omega\rangle,\mathcal{O}^{\dagger}(t^{\prime})\mathcal{O}(t+i\beta)|\Omega\rangle)\nonumber\\
  &=\langle\Omega|\mathcal{O}^{\dagger}(t^{\prime})\mathcal{O}(t+i\beta)|\Omega\rangle,\label{eq:16}
\end{align}
where the second line follows from the assumption \eqref{eq:15}, the
third line the antiunitarity of $J$ (i.e.,
$(J|\Psi\rangle,J|\Phi\rangle)=\overline{(|\Psi\rangle,|\Phi\rangle)}=(|\Phi\rangle,|\Psi\rangle)$),
and the fifth line the relations
$\mathrm{e}^{-\beta H}\mathcal{O}(t)\mathrm{e}^{\beta
  H}=\mathcal{O}(t+i\beta)$ and
$\mathrm{e}^{-\beta H}|\Omega\rangle=|\Omega\rangle$. Setting
$t^{\prime}=0$ we get the condition
$G^{+}(t)=G^{-}(t+i\beta)$. Likewise, one can prove
$G^{-}(t)=G^{+}(t-i\beta)$. These mean that, if \eqref{eq:15} holds,
the Wightman functions with respect to the state $|\Omega\rangle$ are
nothing but the thermal Wightman functions at temperature $T=1/\beta$
(except the question of analyticity on the strips).

The above discussion, though simplified, captures the essence of the
interplay between the KMS condition and the Bisognano--Wichmann
theorem \cite{Bisognano:1975ih,Bisognano:1976za}. In the mid-1970s
Bisognano and Wichmann showed that there exists the identity
\eqref{eq:15} in generic Poincar\'{e}-invariant quantum field
theories. There, the state $|\Omega\rangle$ is the vacuum state
$|0\rangle$ for inertial observers, $J$ is the CPT conjugate (with a
partial reflection), and $\tfrac{\beta}{2\pi}H$ is the generator of
Lorentz boost. The temporal coordinate $t$ is proportional to the
dimensionless Lorentz boost parameter $\theta$ and identified as
$\theta=(2\pi/\beta)t$. Physically speaking, $t$ is identical to the
proper time for uniformly accelerating observers and the proportional
coefficient $2\pi/\beta$ is identical to the proper acceleration $a$,
from which we can deduce the Unruh temperature $T=a/(2\pi)$. This is
the physical content of Bisognano--Wichmann theorem, which provides a
nonperturbative proof for the thermality of Wightman functions with
respect to the vacuum \cite{Sewell:1982zz}.

Now we have come to the point. From a group theoretical viewpoint the
most important thing in the Bisognano--Wichmann theorem is that the
time-translation generator $H$ is given by the Lorentz boost
generator---the generator of one-parameter subgroup $SO(1,1)$ of the
Poincar\'{e} group $ISO(1,d-1)$. In Poincar\'{e}-invariant quantum
field theories the Lorentz boost is the only way to realize the group
$SO(1,1)$ as a coordinate transformation. However, there emerge
several options if the theory enjoys conformal invariance. Typical
examples are the following \cite{Ohya:2016gto}:
\begin{align}
  &SO(1,1): x^{\mu}\mapsto x^{\mu}(\theta)={\Lambda^{\mu}}_{\nu}x^{\nu},\label{eq:17}\\
  &SO(1,1): x^{\mu}\mapsto x^{\mu}(\theta)=\mathrm{e}^{-\theta}x^{\mu},\label{eq:18}\\
  &SO(1,1): x^{\mu}\mapsto x^{\mu}(\theta)=\mathrm{e}^{-\varphi}\frac{x^{\mu}-b^{\mu}(x\cdot x)}{1-2(b\cdot x)+(b\cdot b)(x\cdot x)}+a^{\mu},\label{eq:19}
\end{align}
where
$\Lambda=\left(\begin{smallmatrix}\cosh\theta&\sinh\theta&\\\sinh\theta&\cosh\theta&\\&&\vec{1}_{d-2}\\\end{smallmatrix}\right)$,
$\varphi=2\log\cosh\tfrac{\theta}{2}$,
$b^{\mu}=(\tfrac{1}{\ell}\tanh\tfrac{\theta}{2},0,\cdots,0)$, and
$a^{\mu}=(\ell\tanh\tfrac{\theta}{2},0,\cdots,0)$. Note that
\eqref{eq:17} is a Lorentz boost on the $(x^{0},x^{1})$-plane,
\eqref{eq:18} is a dilatation, and \eqref{eq:19} is a special
conformal transformation followed by a dilatation followed by a
translation. Note also that these transformations are the solutions to
the following flow equations generated by the conformal Killing
vectors $k^{\mu10}$, $k^{\mu d,d+1}$, and $k^{\mu d0}=-k^{\mu0d}$:
\begin{align}
  \Dot{x}^{\mu}(\theta)&=\eta^{\mu1}x^{0}(\theta)-\eta^{\mu0}x^{1}(\theta),\label{eq:20}\\
  \Dot{x}^{\mu}(\theta)&=-x^{\mu}(\theta),\label{eq:21}\\
  \Dot{x}^{\mu}(\theta)&=-\frac{\ell^{2}-x(\theta)\cdot x(\theta)}{2\ell}\eta^{\mu0}-\frac{x^{\mu}(\theta)x^{0}(\theta)}{\ell},\label{eq:22}
\end{align}
where dot stands for the derivative with respect to $\theta$.

Now we wish to identify the parameter $\theta$ with the temporal
coordinate $t$ (up to the factor $2\pi/\beta$). To justify this, the
above conformal Killing vectors must be \textit{timelike}; that is,
$\Dot{x}(\theta)\cdot\Dot{x}(\theta)<0$. It is a straightforward
exercise to classify their timelike domains. The results are as
follows (see also Figure \ref{fig:1}):
\begin{itemize}
\item\textbf{Rindler wedge.} The Killing vector \eqref{eq:20} becomes
  timelike in the following domains:
  \begin{equation}
    W_{\pm}=\{x^{\mu}: \pm x^{1}>|x^{0}|\},\label{eq:23}
  \end{equation}
  which are nothing but the right and left Rindler wedges. The
  coordinate systems in which \eqref{eq:17} yields the
  time-translation are given by
  \begin{equation}
    x^{0}=\pm\ell\frac{\sinh(t/\ell)}{H^{0}+H^{1}},\quad
    x^{1}=\pm\ell\frac{\cosh(t/\ell)}{H^{0}+H^{1}},\quad
    x^{i}=\ell\frac{H^{i}}{H^{0}+H^{1}},\label{eq:24}
  \end{equation}
  where $H^{\mu}=(H^{0},H^{1},\cdots,H^{d-1})$ describes the upper
  half of two-sheeted hyperbolic space $\mathbb{H}^{d-1}$ and is
  subject to the conditions
  $H\cdot H\equiv-(H^{0})^{2}+(H^{1})^{2}+\cdots+(H^{d-1})^{2}=-1$ and
  $H^{0}\geq1$. The induced metrics on $W_{\pm}$ are
  \begin{equation}
    ds_{W_{\pm}}^{2}=\frac{-dt^{2}+\ell^{2}dH\cdot dH}{(H^{0}+H^{1})^{2}}.\label{eq:25}
  \end{equation}
\item\textbf{Light-cone.} The conformal Killing vector \eqref{eq:21}
  becomes timelike in the following domains:
  \begin{equation}
    V_{\pm}=\{x^{\mu}: \pm x^{0}>|\vec{x}|\},\label{eq:26}
  \end{equation}
  which are nothing but the future and past light-cones. The
  coordinate systems in which \eqref{eq:18} yields the
  time-translation are given by
  \begin{equation}
    x^{\mu}=\pm\ell\mathrm{e}^{-t/\ell}H^{\mu},\label{eq:27}
  \end{equation}
  where $H^{\mu}$ is the same as above. The induced metrics on
  $V_{\pm}$ are
  \begin{equation}
    ds_{V_{\pm}}^{2}=\mathrm{e}^{-2t/\ell}(-dt^{2}+\ell^{2}dH\cdot dH).\label{eq:28}
  \end{equation}
\item\textbf{Diamond.} The conformal Killing vector \eqref{eq:22}
  becomes timelike in the following domain:\footnote{In fact, as
    depicted in Figure \ref{fig:1} the conformal Killing vector
    \eqref{eq:22} becomes timelike also in the domains
    $K=\{x^{\mu}: |\vec{x}|-|x^{0}|>\ell\}$ and
    $\mathcal{V}_{\pm}=\{x^{\mu}: \pm x^{0}>|\vec{x}|+\ell\}$.}
  \begin{equation}
    D=\{x^{\mu}: |\vec{x}|+|x^{0}|<\ell\},\label{eq:29}
  \end{equation}
  which is nothing but the diamond (or double cone). The coordinate
  system in which \eqref{eq:19} yields the time-translation is given
  by
  \begin{equation}
    x^{0}=\ell\frac{\sinh(t/\ell)}{\cosh(t/\ell)+H^{0}},\quad
    x^{i}=\ell\frac{H^{i}}{\cosh(t/\ell)+H^{0}},\label{eq:30}
  \end{equation}
  where $H^{\mu}$ is the same as above. The induced metric on $D$ is
  \begin{equation}
    ds_{D}^{2}=\frac{-dt^{2}+\ell^{2}dH\cdot dH}{(\cosh(t/\ell)+H^{0})^{2}}.\label{eq:31}
  \end{equation}
\end{itemize}
Now it is obvious that these subspaces of the flat Minkowski spacetime
$\mathbb{R}^{1,d-1}$ are all conformal to
$\mathbb{H}^{1}\times\mathbb{H}^{d-1}\ni(t,H^{\mu})$. Hence the
correlation functions on $\mathbb{H}^{1}\times\mathbb{H}^{d-1}$ with
respect to the inertial vacuum $|0\rangle$ are just given by conformal
transformations of those in the Cartesian coordinate system. For
example, the positive- and negative-frequency two-point Wightman
functions
$\langle0|\mathcal{O}_{\Delta}(t,H)\mathcal{O}_{\Delta}^{\dagger}(t^{\prime},H^{\prime})|0\rangle$
and
$\langle0|\mathcal{O}_{\Delta}^{\dagger}(t^{\prime},H^{\prime})\mathcal{O}_{\Delta}(t,H)|0\rangle$
are given by
\begin{equation}
  \left[\frac{2\pi^{2}T^{2}}{-\cosh(2\pi T(t-t^{\prime}\mp i\epsilon))-H\cdot H^{\prime}}\right]^{\Delta},\label{eq:32}
\end{equation}
where $T=1/(2\pi\ell)$. It can be shown that these Wightman functions
indeed satisfy the KMS condition and hence give the thermal
correlation functions on $\mathbb{H}^{1}\times\mathbb{H}^{d-1}$ at
temperature $T$ \cite{Ohya:2016gto}. We note that there also exist
theorems \cite{Buchholz:1978,Hislop:1981uh} which generalize the
Bisognano--Wichmann theorem and consider the conformal Killing vectors
\eqref{eq:21} and \eqref{eq:22} and their timelike domains.

So far we have considered correlation functions in position space. For
practical applications, however, we often need to know correlation
functions in momentum space. A standard approach to momentum-space
correlators is the Fourier transform of position-space
correlators. However, the Fourier transform of correlation functions
is generally hard to carry out. In fact, the Fourier transform of
\eqref{eq:32} is, though not impossible, quite complicated and
requires a lot of integration techniques. Hence it would be desirable
to develop a method which bypasses Fourier integrals and directly
leads to momentum-space expressions.\footnote{In zero-temperature CFT
  such a method would simply fall into the study of conformal
  Ward--Takahashi identities in momentum space. In Euclidean signature
  this approach was thoroughly discussed in \cite{Bzowski:2013sza}
  (see also \cite{Coriano:2013jba}).} In the rest of the paper we will
see that the intertwining relations do the job: The operator
identities \eqref{eq:12} enable us to deduce momentum-space two-point
functions in a purely Lie-algebraic fashion.

\section{Intertwining Relations in the $SO(1,1)$ Basis}
\label{sec:4}
Let us finally move on to the intertwining relations in the $SO(1,1)$
diagonal basis---the conformal Ward--Takahashi identities at finite
temperature. We emphasize that this section is rather sketchy. For
more details we refer to our paper \cite{Ohya:2016gto}. In what
follows we shall set $2\pi T=1/\ell=1$ for simplicity. The temperature
dependence is easily restored by dimensional analysis.

To begin with, let us consider the quadratic Casimir operator of the
Lie algebra $\mathfrak{so}(2,d)$, which is given by
\begin{equation}
  C_{2}[\mathfrak{so}(2,d)]=\frac{1}{2}J_{ab}J^{ab}.\label{eq:33}
\end{equation}
We wish to identify the $SO(1,1)$ generator as the time-translation
generator $H$. In group theoretical language, this means that we need
to work with the basis where the following subgroup becomes diagonal:
\begin{equation}
  SO(1,1)\times SO(1,d-1)\subset SO(2,d).\label{eq:34}
\end{equation}
Correspondingly, the quadratic Casimir operator is decomposed as
follows:
\begin{equation}
  C_{2}[\mathfrak{so}(2,d)]=-H(H\pm id)-\eta_{ab}E^{\mp a}E^{\pm b}+C_{2}[\mathfrak{so}(1,d-1)],\label{eq:35}
\end{equation}
where $E^{\pm a}$ are certain linear combinations of $J^{ab}$ and
$C_{2}[\mathfrak{so}(1,d-1)]$ is the quadratic Casimir operator of the
subalgebra $\mathfrak{so}(1,d-1)$. For example, in the case of Rindler
wedge we have $H=J^{10}$, $E^{\pm a}=J^{0a}\pm J^{1a}$, and
$C_{2}[\mathfrak{so}(1,d-1)]=(1/2)J_{ab}J^{ab}$, where $a$ and $b$ run
through $2$ to $d+1$.

Now let $|\Delta,\omega,k;\sigma\rangle$ be a simultaneous eigenstate
of $C_{2}[\mathfrak{so}(2,d)]$, $H$, and $C_{2}[\mathfrak{so}(1,d-1)]$
that satisfies the following eigenvalue equations:
\begin{align}
  C_{2}[\mathfrak{so}(2,d)]|\Delta,\omega,j;\sigma\rangle&=\Delta(\Delta-d)|\Delta,\omega,j;\sigma\rangle,\label{eq:36}\\
  H|\Delta,\omega,j;\sigma\rangle&=\omega|\Delta,\omega,j;\sigma\rangle,\label{eq:37}\\
  C_{2}[\mathfrak{so}(1,d-1)]|\Delta,\omega,j;\sigma\rangle&=j(j-d+2)|\Delta,\omega,j;\sigma\rangle,\label{eq:38}
\end{align}
where $\sigma$ stands for eigenvalues of other simultaneously
commuting generators which are irrelevant in the following
discussion. Below we shall focus on the case $j(j-d+2)<-(d-2)^{2}/4$
and parameterize $j$ as follows:
\begin{equation}
  j=\frac{d-2}{2}+ik,\quad k\in(0,\infty).\label{eq:39}
\end{equation}
In other words, we shall focus on the principal series representation
of $\mathfrak{so}(1,d-1)$. Note that $j(j-d+2)=-k^{2}-(d-2)^{2}/4$ is
real though $j$ is complex. Physically, $k$ plays the role of the
modulus of spatial momentum. From now on we shall write the eigenstate
as $|\Delta,\omega,k;\sigma\rangle$.

Now there are two important things for the following discussion. The
first is that the eigenvalue $\Delta(\Delta-d)$ is invariant under the
exchange $\Delta\to d-\Delta$, which means that the vectors
$|\Delta,\omega,k;\sigma\rangle$ and
$|d-\Delta,\omega,k;\sigma\rangle$ share the same eigenvalue of
$C_{2}[\mathfrak{so}(2,d)]$. These two vectors are mapped to each
other by the intertwining operators and satisfy the following
relations:
\begin{equation}
  G_{\alpha}|d-\alpha,\omega,k;\sigma\rangle=\Tilde{G}_{\alpha}(\omega,k)|\alpha,\omega,k;\sigma\rangle,\quad\alpha\in\{\Delta,d-\Delta\},\label{eq:40}
\end{equation}
where the proportional coefficients $\Tilde{G}_{\alpha}(\omega,k)$ are
the momentum-space two-point functions. From now on $J_{\alpha}^{ab}$,
$H_{\alpha}$, $E_{\alpha}^{\pm a}$, etc. denote the $SO(2,d)$
generators that act on the representation space spanned by the vectors
$\{|\alpha,\omega,k;\sigma\rangle\}$. For example, their differential
representations are given in \eqref{eq:7}.

The second important thing is the set of generators
$E_{\alpha}^{\pm a}$. One can show that there exist certain linear
combinations $E_{\alpha}^{\pm}$ of these generators that satisfy the
following ladder equations:
\begin{align}
  E^{\pm}_{\alpha}|\alpha,\omega,k;\sigma\rangle
  &=A^{\pm}\left[\alpha-\tfrac{d-2}{2}\mp i(\omega\pm k)\right]|\alpha,\omega\pm i,k+i;\sigma\rangle\nonumber\\
  &\quad+B^{\pm}\left[\alpha-\tfrac{d-2}{2}\mp i(\omega\mp k)\right]|\alpha,\omega\pm i,k-i;\sigma\rangle.\label{eq:41}
\end{align}
For example, in the case of Rindler wedge they are given by
$E_{\alpha}^{\pm}=E_{\alpha}^{\pm d}+E_{\alpha}^{\pm(d+1)}$. Note that
$A^{\pm}$ and $B^{\pm}$ are $\alpha$-independent irrelevant factors.

Now we have almost done. Let us finally consider the intertwining
relations
$E^{\pm}_{\Delta}G_{\Delta}=G_{\Delta}E^{\pm}_{d-\Delta}$. Applying
these to the state $|d-\Delta,\omega,k;\sigma\rangle$ we get
\begin{equation}
  E^{\pm}_{\Delta}G_{\Delta}|d-\Delta,\omega,k;\sigma\rangle=G_{\Delta}E^{\pm}_{d-\Delta}|d-\Delta,\omega,k;\sigma\rangle.\label{eq:42}
\end{equation}
It follows from \eqref{eq:40} and \eqref{eq:41} that the identities
\eqref{eq:42} result in the following nontrivial functional equations
in complex momentum space:
\begin{align}
  \Tilde{G}_{\Delta}(\omega\pm i,k\pm i)&=\frac{\Delta-\frac{d-2}{2}\mp i(\omega+k)}{\Tilde{\Delta}-\frac{d-2}{2}\mp i(\omega+k)}\Tilde{G}_{\Delta}(\omega,k),\label{eq:43}\\
  \Tilde{G}_{\Delta}(\omega\pm i,k\mp i)&=\frac{\Delta-\frac{d-2}{2}\mp i(\omega-k)}{\Tilde{\Delta}-\frac{d-2}{2}\mp i(\omega-k)}\Tilde{G}_{\Delta}(\omega,k),\label{eq:44}
\end{align}
where $\Tilde{\Delta}=d-\Delta$ is the scaling dimension of the shadow
operator. Since these are kind of recurrence relations, we can guess
the solution by iteration. ``Minimal'' solutions to the recurrence
relations are as follows:
\begin{equation}
  \Tilde{G}_{\Delta}^{\pm}(\omega,k)
  \propto\mathrm{e}^{\pm\pi\omega}
  \left|\Gamma\left(\tfrac{\Delta-\frac{d-2}{2}+i(\omega+k)}{2}\right)\right|^{2}
  \left|\Gamma\left(\tfrac{\Delta-\frac{d-2}{2}+i(\omega-k)}{2}\right)\right|^{2},\label{eq:45}
\end{equation}
which can be interpreted as the positive- and negative-frequency
Wightman functions. Indeed, these satisfy the KMS condition in
momentum space,
$\Tilde{G}_{\Delta}^{+}(\omega,k)=\mathrm{e}^{2\pi\omega}\Tilde{G}_{\Delta}^{-}(\omega,k)$.
One can also check that the solutions \eqref{eq:45} exactly coincide
with the Fourier transform of \eqref{eq:32} \cite{Ohya:2016gto}. Note
that $T$ can be restored by the replacements
$\omega\to\omega/(2\pi T)$ and $k\to k/(2\pi T)$.

To summarize, we have seen that the intertwining relations, which are
just the conformal Ward--Takahashi identities in disguise, result in
the recurrence relations \eqref{eq:43} and \eqref{eq:44} when applied
to the $SO(1,1)$ diagonal basis. These are the conformal
Ward--Takahashi identities at finite temperature and give us
nontrivial constraints on momentum-space two-point functions. Though
may need a bit of experience, one can deduce the momentum-space
correlators from these constraints without recourse to the notoriously
complicated Fourier transform. We think this is a big step toward the
understanding of real-time momentum-space correlators in
$d(\geq3)$-dimensional finite-temperature CFT, because these have not
been studied in the literature. In fact, for $d\geq3$ and at nonzero
temperature, even the momentum-space two-point functions of scalar
primary operators have been unknown. It would be quite interesting to
generalize our approach to thermal spinning correlators.

\end{document}